\documentclass[aps,pra,reprint,floatfix,longbibliography]{revtex4-2}
\usepackage{amsmath,amssymb}
\usepackage{graphicx,booktabs}

\newcommand{\Leff}{\Lambda_{\rm eff}}
\newcommand{\reff}{\rho_{\rm eff}}
\newcommand{\tmin}{t_{\min}}

\newcommand{\kB}{k_{B}}

\begin{document}

\title{Magnetic susceptibility and superfluid density of\\ inhomogeneous superconducting films: the $T_c$--mosaic model}
\author{Ruslan Prozorov}
\email{prozorov@ameslab.gov}
\affiliation{Ames National Laboratory, Ames, Iowa 50011, USA}
\affiliation{Department of Physics and Astronomy, Iowa State University, Ames, Iowa 50011, USA}
\date{25 October 2026}

\begin{abstract}
Whole-sample and mutual-inductance part-sample measurements of the linear magnetic susceptibility, $\chi$, of superconducting films are routinely used to obtain the London penetration depth, $\lambda(T)$, assuming uniform superconductivity. However, spatially resolved probes show that many films of current interest are inhomogeneous on micrometer scales. Here we ask whether a distribution of local transition temperatures in a fully gapped film can produce an apparent line-nodal, $T$-linear $\lambda(T)$. The film is modeled as a random Voronoi mosaic of domains with $T_c$ distributed between $T_{c,\min}$ and $T_{c,\max}$ with s-wave (also d-wave, for comparison) London response inside each domain. We show that detectable screening appears only below the percolation temperature, $T_p$, and the transition broadens with the spread of $T_c$. For s-wave domains, $\lambda(T)$ remains activated for any $T_{c,\min}>0$. Only for $T_{c,\min}=0$ does it become a power law, $\Delta\lambda\propto T^n$ with $n\approx1.7$, but then the onset of screening drops to $T_p=T_{c,\max}/2$. Weighting the $T_{c}$ distribution towards $T_c\to0$ (more low-$T_{c}$ domains) can bring $n$ closer to 1 ($n\approx1.2$), but simultaneously pushes the apparent transition, $T_p$, more than an order of magnitude below $T_{c,\max}$. For d-wave domains, the linear term survives, and its relative slope is enhanced almost fourfold for $T_{c,\min}=0$.

Hence, a $T$-linear $\lambda(T)$ measured in a film with screening onset within a factor of a few of $T_{c,\max}$ by whole-sample or mutual-inductance part-sample measurements cannot originate from $T_c$ mosaicity in a fully gapped superconductor.
\end{abstract}

\maketitle

\section{Introduction}
The temperature-dependent London penetration depth, $\lambda(T)$, is a primary probe of the superconducting gap structure \cite{prozorov2006,prozorov2011a}. It is usually obtained from the small-amplitude, vortex-free linear magnetic susceptibility, $\chi(T)$. The susceptibility is measured on the whole sample (e.g., using conventional amplitude-domain AC susceptibility \cite{topping2019} and frequency-domain tunnel-diode resonator \cite{giannetta2022a,moirangthem2026b}) or using a mutual inductance technique, where a macroscopic area of the sample is exposed to a localized magnetic field \cite{fiory1988,turneaure1998,yan2026a}. Importantly, in thin films, the extraction of $\lambda(T)$ from $\chi(T)$ requires knowing $\chi(\lambda)$ and assumes a homogeneous superconducting state \cite{prozorov2026films,moirangthem2026b}.

The extracted $\lambda(T)$ is then used to evaluate the superfluid density, $\rho_s=(\lambda(0)/\lambda(T))^2$ (or proportional to it, the superconducting phase stiffness). In some cases, its low-temperature behavior is distinctly non-activated. For example, a $T-$linear $\lambda(T)$ has recently been observed in FeTe-based heterostructures \cite{moirangthem2026b}. A closely related local scanning SQUID measurement of FeTe films shows a similar behavior inside the regions of stronger superconductivity \cite{li2026fete}. Both suggest a line-nodal order parameter.
However, the superconducting state in real films, especially those with complex stoichiometry and fabrication protocols (including the above-mentioned FeTe), is often inhomogeneous, as detected by spatially resolved local measurements on scales of a few to tens of micrometers \cite{iguchi2001,ferguson2024,yi2025a,li2026fete,li2026a}.

Here we ask whether the magnetic susceptibility of a spatially inhomogeneous s-wave film can yield a superfluid density consistent with apparent line-nodal behavior. For comparison, we also study what happens if the film is intrinsically d-wave. Does $\chi(T)$ retain a sharp feature at the transition to the normal state? How does $\chi(T)$ shape change over the full temperature range? What happens with the low-temperature behavior of $\rho_s(T)$? To answer these questions, we combine sheet-current London electrodynamics with the percolation theory of random two-dimensional media \cite{kirkpatrick1973,efros1976,stauffer1994}. In that theory, the superfluid density of a superconductor--insulator mixture maps onto the conductance of a conductor--insulator mixture. We computed the macroscopic magnetic susceptibility for different transition-temperature distributions and analyzed the temperature dependence of the superfluid density and London penetration depth.

\section{Model}
\subsection{The $T_c$--mosaic}
The film is tiled by a Poisson--Voronoi tessellation \cite{okabe2000}. The seeds are placed randomly with areal density $4/(\pi\ell^2)$, and each point in the film belongs to the domain of the nearest seed. Here $\ell$ is the diameter of a disc whose area equals the mean area of all mosaic domains, $\pi\ell^2/4$. Domains have random sizes and shapes. We used $\ell=20\,\mu$m as the characteristic size suggested by the experiment. The domain $i$ has its own transition temperature $T_{c,i}$, selected independently and uniformly from the interval $[T_{c,\min},T_{c,\max}]$ (other distributions are considered in Sec.~\ref{sec:nonuni}). All temperatures are normalized by $T_{c,\max}$: $t=T/T_{c,\max}$, $t_i=T_{c,i}/T_{c,\max}$, $t_p=T_p/T_{c,\max}$ and $\tmin=T_{c,\min}/T_{c,\max}$, $t_{\max}=1$.
Domain boundaries are sharp. Adjacent superconducting domains share one superconducting condensate, and the supercurrent passes freely between them. A normal domain carries no supercurrent and provides no proximity or Josephson coupling across it.

\begin{figure*}[t]
\includegraphics[width=17cm]{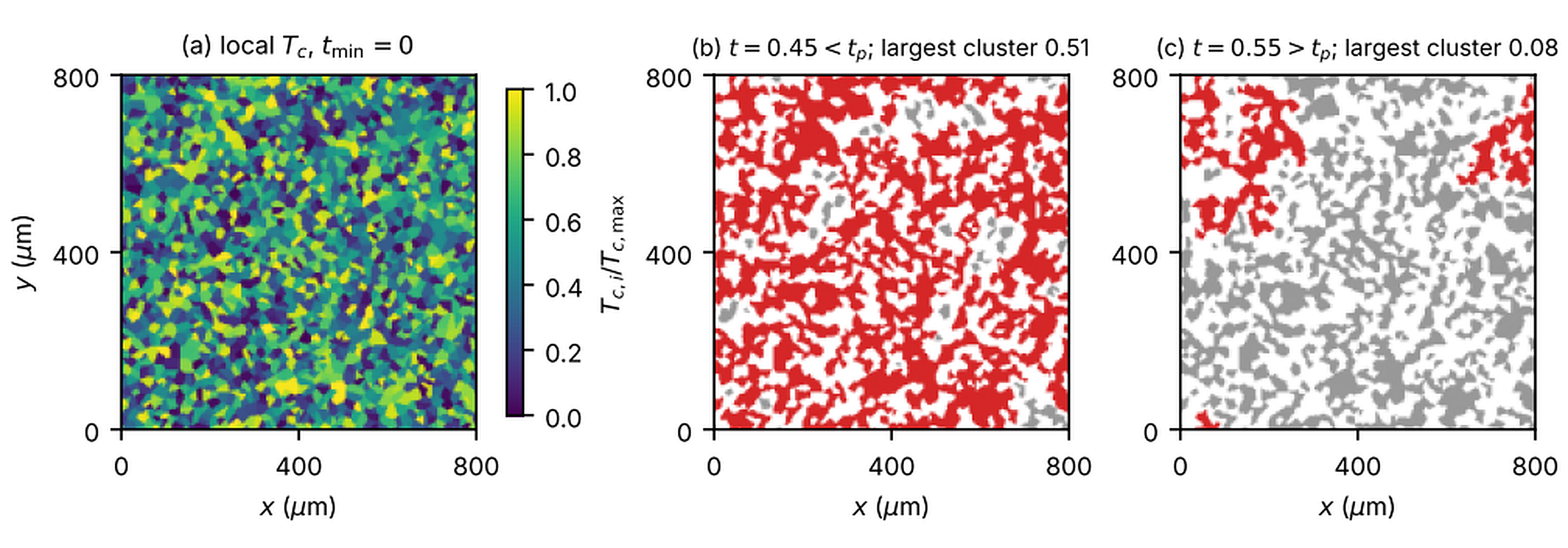}
\caption{(a) An $800\times800\,\mu$m$^2$ realization of the mosaic at $\tmin=0$ with mean domain diameter $\ell=20\,\mu$m colored by the local $T_c$. (b,c) Superconducting domains ($t_i>t$) just below and just above the percolation temperature $t_p(\tmin=0)=0.5$. The largest connected cluster is shown in red. Below $t_p$ it spans the box (periodic boundaries, percolation); above $t_p$ only finite non-percolating clusters remain.}
\label{fig:mosaic}
\end{figure*}

Figure~\ref{fig:mosaic}(a) shows an example of a mosaic realization with the broadest distribution, $\tmin=0$, where the domains are colored by their $T_c$. Panels (b) and (c) show the superconducting domains below, $t=0.45$, and above, $t=0.55$, the percolation threshold, which in this case is $t_p(\tmin=0)=0.5$.

\subsection{Local superfluid density}

Within domain $i$ the London penetration depth $\lambda_i(T)$ and the Pearl length $\Lambda_i(T)$ are,
\begin{equation}
\begin{split}
\frac{1}{\lambda_i^2(T)}&=\frac{t_i}{\lambda_1^2(0)}\,\rho\!\left(\frac{T}{T_{c,i}}\right),\\
\Lambda_i(T)&=\frac{2\lambda_i^2(T)}{d}=\frac{\Lambda_1(0)}{t_i\,\rho(T/T_{c,i})}.
\end{split}
\label{eq:lami}
\end{equation}

For our numerical calculation, the zero-temperature London penetration depth, $\lambda(0)$, of a domain with the highest $T_c=T_{c,\max}$ is set to $\lambda_1(0)=1\,\mu$m, of the order of the values reported for FeTe-based films \cite{moirangthem2026b,li2026fete}, so that $\Lambda_1(0)=2\lambda_1^2(0)/d=100\,\mu$m for $d=20$~nm. To avoid extra parameters, for an arbitrary domain $i$ we use $\lambda_i^{-2}(0)\propto \sigma T_{c,i}$ with a common normal-state conductivity $\sigma$. This so-called Homes scaling was reported for a variety of superconductors  \cite{homes2004} and, in fact, follows from the BCS theory with disorder for a wide range of scattering rates \cite{kogan2013e}.

Since the main question is whether some spatial distributions of $T_c$ in an s-wave mosaic film can produce a superfluid density that looks line-nodal ($T$-linear variation at $t\to 0$), we use as intrinsic references for the temperature dependence of the order parameter both a fully gapped isotropic s-wave and a line-nodal d-wave for comparison.

The temperature dependence of the normalized superfluid density, $\rho(t)=\lambda^2(0)/\lambda^2(T)$, is computed self-consistently in the full temperature range in the weak-coupling clean limit for a cylindrical Fermi surface, with the superconducting order parameter $\Delta_\varphi(T)=\Delta(T)\,\eta(\varphi)$, where $\eta=1$ for s-wave and $\eta=\cos2\varphi$ for d-wave, so that $\Delta(T)$ is the maximum gap in both cases. At each temperature, the amplitude $\Delta(T)$ is obtained by solving the self-consistency (gap) equation,
\begin{equation}
\ln\frac{1}{t}=\frac{2\pi\kB T}{\langle\eta^2\rangle_\varphi}\sum_{n\ge0}\left\langle\eta^2(\varphi)\left[\frac{1}{\omega_n}-\frac{1}{\sqrt{\omega_n^2+\Delta_\varphi^2}}\right]\right\rangle_{\!\varphi},
\label{eq:gap}
\end{equation}
where $\omega_n=\pi\kB T(2n+1)$ are the Matsubara frequencies and $t=T/T_c$. The superfluid density follows from the Matsubara sum,
\begin{equation}
\rho(t)=2\pi\kB T\sum_{n\ge0}\left\langle\frac{\Delta_\varphi^2}{(\omega_n^2+\Delta_\varphi^2)^{3/2}}\right\rangle_{\!\varphi}.
\label{eq:rho}
\end{equation}
There are no adjustable parameters: $T_c$ fixes $\Delta(T)$ and $\rho(t)$.

For an s-wave order parameter, $\Delta(0)=1.764\,\kB T_c$, and $\rho\simeq 1-\sqrt{2\pi\Delta(0)/\kB T}\,e^{-\Delta(0)/\kB T}$ is exponentially small at low temperatures. The dirty-limit s-wave form, $\rho=(\Delta(T)/\Delta(0))\tanh{(\Delta(T)/2\kB T)}$, would differ from Eq.~(\ref{eq:rho}) by at most $0.09$ (at $t\approx0.7$), and it has the same exponential low-temperature behavior \cite{tinkham1996}.

For a d-wave order parameter, $\Delta(0)=2.14\,\kB T_c$, and $\rho\simeq1-(2\ln2)\kB T/\Delta(0)=1-0.648\,T/T_c$ at low temperature, i.e.,  $\Delta\lambda/\lambda(0)=0.324\,T/T_c$ \cite{prozorov2006}.
The impurity rounding of the d-wave linear term, $T\to T^2/(T+T^*)$, is not included. This formula was derived for strong (unitary) limit scattering to explain the experiments in high-$T_c$ cuprates \cite{hirschfeld1993}. In a more common weak-scattering (Born) limit, the $T \to T^2$ rounding becomes noticeable only above a very large scattering rate, $\Gamma \gtrapprox 0.17$, where $\Gamma=\hbar/(2\pi\kB T_{c0}\tau)$, $\tau$ is the total (transport plus spin-flip) Born scattering time, and $T_{c0}$ is the clean-limit $T_c$, when the transition temperature is almost half that of the clean state, $T_c(\Gamma=0.17)=0.54T_c(\Gamma=0)$. For reference, the critical value at which d-wave superconductivity is quenched by non-spin-flip scattering is $T_c(\Gamma_{\rm crit} = 0.28)=0$ \cite{kogan2013c}.

Each domain enters the calculation only through what we call its ``contrast" relative to the strongest domain,
\begin{equation}
c_i(T)\equiv\frac{\Lambda_i(T)}{\Lambda_1(T)}=\begin{cases}\dfrac{\rho(t)}{t_i\,\rho(t/t_i)}, & t<t_i,\\[4pt] \infty, & t\ge t_i.\end{cases}
\label{eq:contrast}
\end{equation}

\subsection{Sheet currents in the $T_c$--mosaic and homogenization}\label{sec:homog}

Even in an ultra-thin film in a perpendicular magnetic field, the supercurrent density is far from uniform throughout the thickness \cite{prozorov2026films}. However, for the magnetic fields outside the film, and therefore for the magnetic moment, $m$, and magnetic susceptibility, $\chi=m/(HV)$ ($V$ is the sample volume), only the thickness-integrated sheet current, $\mathbf I$ (current per unit width), matters. With Pearl length $\Lambda=2\lambda^2/d$, this 2D sheet description reproduces full three-dimensional solutions for $\lambda\gtrsim4d$ and $d/2a\lesssim10^{-4}$ \cite{prozorov2026films}.

The London equation of the sheet reads,
\begin{equation}
\mathbf I=-\frac{2}{\mu_0\Lambda}\Big(\mathbf A-\frac{\Phi_0}{2\pi}\nabla\theta\Big)=\frac{2m_e}{\mu_0e\Lambda}\,\mathbf v_s,
\label{eq:I}
\end{equation}
where $m_e$ and $e$ are the electron mass and charge, $\Phi_0$ is the flux quantum. Here, $\theta$ is the phase of the order parameter, and $\mathbf v_s$ is the superfluid velocity, both averaged over the thickness. The kinetic energy per unit area is $\mu_0\Lambda I^2/4$. Since $I$ is the actual current and $\rho_s\propto 1/\lambda^2 \propto 1/\Lambda$, Eq.~(\ref{eq:I}) can be thought of as Ohm's law with $\rho_s$ as the scalar ``conductivity'' and $\mathbf v_s$ as the vector ``electric field''. A random network of domains with different $\rho_s$ is therefore the superconducting analog of a random resistor network.

The problem has three well-separated length scales: the microscopic ones ($\xi_i$, $\lambda_i$, $d$), the mesoscopic domain size $\ell=20\,\mu\textrm{m}$, and the macroscopic sample lateral size $a=3\,\textrm{mm}$. The coherence length, $\xi_i$, plays no role, since we solve the London equation. In the calculations, we make the following assumptions:
\begin{enumerate}
\item \emph{Local London response.} $\xi\ll\ell$, so each domain obeys Eq.~(\ref{eq:I}) with its own $\Lambda_i(T)$. Domain walls are sharp on the scale $\ell$. The proximity length in a normal domain, $\sim\sqrt{\hbar D/2\pi\kB T}\lesssim0.1\,\mu$m, is negligible.
\item \emph{Thin-sheet limit.} $d\ll\ell$, and the sheet conditions are met: $\lambda\ge1\,\mu$m $=50d$ and $d/2a=3.3\times10^{-6}$. Only the thickness-integrated current enters, both in the local physics and in the magnetic fields outside the film.
\item \emph{Vortex-free linear response.} The excitation magnetic field amplitude is small and no vortices enter. The fluxoid (flux plus the contribution of screening currents) around every hole in the superconducting network, i.e., every normal island or cluster, keeps its zero-field value.
\item \emph{Scale separation.} $\ell\ll a$. Locally, the macroscopic Meissner current of the disc looks like a uniform mean current $\bar{\mathbf I}$. On the scale $\ell$ the $T_c$ inhomogeneity only redistributes this current around the domains.
\item \emph{Stray fields of the redistributed current.} The non-uniform part of $\mathbf I$ produces magnetic fields that decay away from the film over a distance $\sim\ell$. Their energy is kept in the local problem but does not affect the macroscopic field.
\end{enumerate}

Because $\nabla\cdot\mathbf I=0$, the current can be written with a stream function, $g(\mathbf r)$, $\mathbf I=\bar{\mathbf I}+\nabla\times(g\hat{\mathbf z})$. Function  $g(\mathbf r)$ obeys Born--von~K\'arm\'an periodic boundary conditions on the 2D $L\times L$ computational box. The mean current is carried by the non-periodic linear part of the stream function.
The energy per unit area of the local problem is
\begin{equation}
\frac{4}{\mu_0}E[g]=\big\langle\Lambda(\mathbf r)\,|\bar{\mathbf I}+\nabla\times g\hat{\mathbf z}|^2\big\rangle+\sum_{\mathbf k\neq0}|\mathbf k|\,|g_{\mathbf k}|^2,
\label{eq:E}
\end{equation}
where $g_{\mathbf k}$ are the Fourier amplitudes of  $g(\mathbf r)$. The first term is the kinetic energy. The second is the energy of the magnetic field produced by the current pattern in the half-spaces above and below the film ($|z|>d/2$. The field energy inside the film is negligible for $d\ll\ell$). For a uniform sheet, the two terms give $|\mathbf k|(1+\Lambda|\mathbf k|)|g_{\mathbf k}|^2$, which is Pearl's kernel \cite{pearl1964}. The minimum of Eq.~(\ref{eq:E}) at fixed $\bar{\mathbf I}$ defines the homogenized Pearl length, $\Leff$,
\begin{equation}
\min_g E=\frac{\mu_0}{4}\Leff\bar I^2,\qquad G\equiv\frac{\Leff}{\Lambda_1(T)},
\label{eq:G}
\end{equation}
where $\Lambda_1(T)$ is the Pearl length of a ``strongest superconductivity'' domain with $T_c=T_{c,\max}$, Eq.~(\ref{eq:lami}) at $t_i=1$. The local Pearl length in Eq.~(\ref{eq:E}) can be  written as $\Lambda(\mathbf r)=\Lambda_1(T)\,c(\mathbf r)$, with what we call a ``contrast" $c(\mathbf r)=c_i$ for $\mathbf r$ in domain $i$ [Eq.~(\ref{eq:contrast})]. The dimensionless ratio is $G=1$ for a uniform film with $T_c=T_{c,\max}$ and $G\ge1$ otherwise.

The Euler--Lagrange equation of Eq.~(\ref{eq:E}) is the London equation for the sheet current $\mathbf I$, with Pearl's $\Lambda$ as the natural length scale,
\begin{equation}
\frac{\mu_0}{2}\,\hat{\mathbf z}\cdot\nabla\times\big[\Lambda(\mathbf r)\mathbf I\big]=-B_z(\mathbf r),
\label{eq:london}
\end{equation}
where $B_z(\mathbf r)$ is the magnetic field produced by the current $\mathbf I$ in the plane of the film. Inside a normal domain, $\Lambda\to\infty$ forces $\mathbf I\to0$, so the stream function, including its linear part, is constant there. This constant is a free variable of the minimization of Eq.~(\ref{eq:E}). Stationarity with respect to this constant sets the change of the fluxoid around the domain to zero, which is assumption 3.

The magnetic term describes the screening of the redistributed currents by their own stray fields. It is small when $\ell\ll\Leff$. Without it, Eq.~(\ref{eq:E}) is the variational principle for the effective conductance of a random conductor with local conductivity $\rho(\mathbf r)\propto1/\Lambda(\mathbf r)$ \cite{kirkpatrick1973}.
Two exact properties of this problem are used below.
\begin{itemize}
\item For a statistically symmetric two-phase random mixture in two dimensions, the Keller--Dykhne duality gives $\reff=\sqrt{\rho_a\rho_b}$ \cite{keller1964,dykhne1971}.
\item Because $\reff$ is a homogeneous first-degree function of the local superfluid densities and the current pattern minimizes $E$, a small change $\delta\rho(\mathbf r)$ changes $\reff$ by
\begin{equation}
\delta\reff=\big\langle\delta\rho(\mathbf r)\,u^2(\mathbf r)\big\rangle,\qquad u(\mathbf r)=\frac{|\mathbf v_s(\mathbf r)|}{|\langle\mathbf v_s\rangle|}.
\label{eq:sens}
\end{equation}

Here $u(\mathbf r)$ is the magnitude of the local superfluid velocity, $\mathbf v_s\propto\Lambda\mathbf I$, Eq.~(\ref{eq:I}), normalized by the magnitude of its average over the film. In the kinetic limit, $u=\Lambda(\mathbf r)|\mathbf I(\mathbf r)|/(\Leff\bar I)$. As in a resistor network, where a poorly conducting inclusion carries little current but a large electric field, a weak domain carries a smaller current but a larger supervelocity: $u<1$ in strong domains and $u>1$ in weak ones. Normal domains do not contribute to Eq.~(\ref{eq:sens}), since $\delta\rho=0$ there.
\end{itemize}

\subsection{Total magnetic susceptibility of a thin disc}

Once $\Leff(T)$ is known, the film is a uniform sheet with Pearl length $\Leff$. It was shown previously that in the sheet-current regime, the total magnetic susceptibility, $\chi$, of a disc of radius $a$ for arbitrary $\lambda$ and aspect ratios $d/2a\le10^{-4}$, is \cite{prozorov2026films}
\begin{equation}
\frac{|\chi|}{\chi_0}=F\!\left(\frac{\Leff}{a}\right),\qquad \chi_0=\frac{8a}{3\pi d},
\label{eq:chi}
\end{equation}
with $\chi_0$ the ideal ($\Lambda\to0$) thin-disc value \cite{clem1994}. The sheet function is (with $\epsilon=\Leff/a$),
\begin{equation}
\frac{1}{F(\epsilon)}-1=\epsilon\Big[\frac{3}{\pi}\ln\Big(1+\frac{c_2}{\epsilon}\Big)+\frac{32}{3\pi}+\frac{c_3}{1+c_4\epsilon}+\frac{c_5\epsilon}{(1+c_6\epsilon)^2}\Big],
\label{eq:F}
\end{equation}
with $c_2=0.0401744$, $c_3=0.8898535$, $c_4=8.093975$, $c_5=20.0621225$ and $c_6=94.44745$ (accuracy $0.02\%$) \cite{prozorov2026films}.
Its limits are
\begin{itemize}
\item $1-F\simeq(3/\pi)\epsilon\ln(3.57/\epsilon)$ for $\epsilon\to0$;
\item $F\simeq(3\pi/32)\epsilon^{-1}$ for $\epsilon\gg1$.
\end{itemize}
For $a=3$~mm the crossover $\Leff=a$ corresponds to $\reff=\Lambda_1(0)/a=0.033$.
Over most of the temperature range the disc is therefore in the strong-screening regime, $\Leff\ll a$, where $|\chi|$ is close to $\chi_0$ and the deficit is small, $1-|\chi|/\chi_0\simeq(3/\pi)(\Leff/a)\ln(3.57\,a/\Leff)$, i.e.\ nearly linear in $\Leff\propto1/\reff$, with a logarithmic correction.

\section{Numerical procedure}\label{sec:numerics}
The effective Pearl length $\Leff(T)$ was computed in six steps. The parameters used are listed in Table~\ref{tab:par}.

\begin{table}[h]
\caption{Parameters used in the calculation.}
\label{tab:par}
\begin{ruledtabular}
\begin{tabular}{ll}
disc radius, thickness & $a=3$~mm, $d=20$~nm\\
$\lambda_1(0)$, $\Lambda_1(0)$ & $1\,\mu$m, $100\,\mu$m\\
domain diameter $\ell$ & $20\,\mu$m, Voronoi\\
local $T_c$ & uniform in $[\tmin,1]\,T_{c,\max}$\\
$\lambda_i^{-2}(0)$ & $\propto T_{c,i}$\\
box, grid & $800\,\mu$m, $256^2$ ($h=3.125\,\mu$m)\\
realizations & 3 (two current directions each)\\
normal-domain contrast & $c=10^6$\\
CG tolerance & $10^{-10}$ (relative residual)
\end{tabular}
\end{ruledtabular}
\end{table}

\emph{Step 1:  $T_c$--mosaic.} A periodic square box of side $L=800\,\mu$m (40 domain diameters) is filled with a Poisson number of seeds, with mean $L^2/(\pi\ell^2/4)\approx2040$. The box is discretized by a square $N\times N$ grid with $N=256$, so the grid spacing (the side of one grid cell) is $h = L/N = 3.125\,\mu\textrm{m}$, and a domain of mean diameter spans 6.4 grid cells. Each grid node $(ih,jh)$ and each edge midpoint, $((i+\tfrac12)h,jh)$ and $(ih,(j+\tfrac12)h)$, is assigned to the domain of its nearest seed, with distances measured periodically across the box edges.

\emph{Step 2: contrast and local superfluid density.} At each temperature, contrast $c_i(t)$ is evaluated from Eq.~(\ref{eq:contrast}). The function $\rho(t)$ of each pairing state is tabulated once on a fine grid by the same code. At each tabulated temperature, Eq.~(\ref{eq:gap}) is solved for $\Delta$ by root bracketing, with the Matsubara sum truncated at 4000 terms plus its analytic tail. Equation~(\ref{eq:rho}) is then summed up to $\omega_n\approx40\Delta$ (at least 2000 terms), again with the analytic tail. The angular average uses 128 points on $[0,\pi/2]$ for d-wave and is trivial for s-wave. Below $t=0.02$ the angular grid cannot resolve the d-wave nodes, so the exact limit shown previously, $1-0.648\,t$, is used. For s-wave, $\rho=1$ to machine precision below $t=0.01$. Normal domains get $c=10^6$, which makes their current negligible (below $10^{-6}$ of the mean).

\emph{Step 3: discretization.} The stream function $g$ is defined on the grid nodes. With periodic forward differences $D_x,D_y$, the current components $I_x=\bar I_x+D_yg$ and $I_y=\bar I_y-D_xg$ live on the edges, where the corresponding contrasts $c^{(y)}$ and $c^{(x)}$ are evaluated. This staggered (Marker-and-Cell, MAC-type) grid conserves current exactly cell by cell. The kinetic energy is $\Lambda_1\langle c^{(y)}I_x^2+c^{(x)}I_y^2\rangle$. The magnetic term is evaluated by fast Fourier transform with the lattice wavenumber $|\mathbf k|_h=(2/h)[\sin^2(k_xh/2)+\sin^2(k_yh/2)]^{1/2}$.

\emph{Step 4: linear system.} Minimizing Eq.~(\ref{eq:E}) gives $[\Lambda_1\mathsf L+\mathsf M]\,g=b$. Here $\mathsf L=D_x^{\rm T}C^{(x)}D_x+D_y^{\rm T}C^{(y)}D_y$ is the kinetic (weighted-Laplacian) operator and $\mathsf M$ is the magnetic operator, diagonal in $\mathbf k$. The right-hand side is $b=-\Lambda_1D_y^{\rm T}c^{(y)}$ for $\bar{\mathbf I}=\hat x$ and $b=+\Lambda_1D_x^{\rm T}c^{(x)}$ for $\bar{\mathbf I}=\hat y$. One node is fixed to remove the constant.

When normal domains are present, the edge contrasts span six orders of magnitude, and conjugate gradients \cite{saad2003} with a diagonal preconditioner fail to converge. For example, for $t_{\min}=0$ at $t=0.4$ they do not reach a required relative residual of $10^{-10}$ in $2\times10^4$ iterations. We therefore factorize $\Lambda_1\mathsf L$ exactly by sparse LU \cite{li2005superlu}. The elimination order controls the fill-in of the factors. With column approximate minimum degree (COLAMD) ordering \cite{davis2004colamd}, the factors contain $1.2\times10^7$ nonzeros (37 times as many as $\mathsf L$) and take about 1 s to compute on one CPU core, against $6.7\times10^7$ non-zeros and 15 s in natural order.

\emph{Step 5: superfluid density.} The energy of the solution gives $G_x$ and $G_y$ through Eq.~(\ref{eq:G}). The values of $1/G$ for three realizations and two directions are averaged, and the standard error of the six values is recorded,
\begin{equation}
\reff(t)=\rho(t)\,\big\langle G^{-1}\big\rangle.
\end{equation}
The kinetic-only value $\rho_{\rm kin}$, obtained from the LU solution before iteration, measures the stray-field correction.

\emph{Step 6: magnetic susceptibility.} $\Leff=\Lambda_1(0)/\reff$ is inserted into Eqs.~(\ref{eq:chi}) and (\ref{eq:F}) with $a=3$~mm.

\emph{Temperature grid.} Each distribution ($\tmin=0.9$, 0.7, 0.5, 0.2 and 0) and pairing state uses 144 temperatures: 100 uniformly spaced in $[0,0.995]$, 12 logarithmically spaced in $[0.005,0.1]$, 30 concentrated around the percolation temperature, and two at $t=0.997$ and 0.999. For the uniform film, $\tmin=1$, the points at $t\ge1$ are dropped, leaving 136.

The numerical calculations were validated using several criteria. The result expected for a uniform film, $G=1$, was reproduced to machine precision. The Keller--Dykhne duality was checked for a 50/50 two-phase Voronoi mosaic with contrast 10 (kinetic only), where the solver gives $G=3.142$ (average over three realizations and two directions), against the exact $\sqrt{10}=3.162$. For contrast 100, it gives $10.03$ against 10. For $\tmin=0$, grid refinement to $512^2$ changes $G$ by less than $0.3\%$ away from $t_p$, and doubling the box to $1600\,\mu$m ($\approx8000$ domains) leaves $\reff$ unchanged within the realization scatter for $t\le0.45$ and sharpens the approach to $t_p$. Including the magnetic term changes $\reff$ by at most $1.3\%$ at $t=0$ and by at most $6\%$ close to $t_p$, where normal clusters are large.

\section{Results}

\subsection{The effective superfluid density}

The domains that are superconducting at temperature $t$ occupy a random fraction $x(t)=(1-t)/(1-\tmin)$ of the film. Site percolation on the Poisson--Voronoi tessellation has a threshold of exactly $1/2$ \cite{bollobas2006}. A macroscopic superfluid density therefore exists only below
\begin{equation}
t_p=\frac{1+\tmin}{2},
\end{equation}
which is the median local $T_c$. Between $t_p$ and 1 the film contains only finite superconducting clusters [Fig.~\ref{fig:mosaic}(c)]. Each cluster screens on its own, but a cluster of size $R\ll\Lambda$ has moment $\sim R^4H/\Lambda$. Their total relative to a connected film is of order $R^2/(a\Lambda)\lesssim10^{-3}$ for $R\sim\ell$, which is invisible on the scale of the figures.

\begin{figure}[tbh]
\includegraphics[width=8cm]{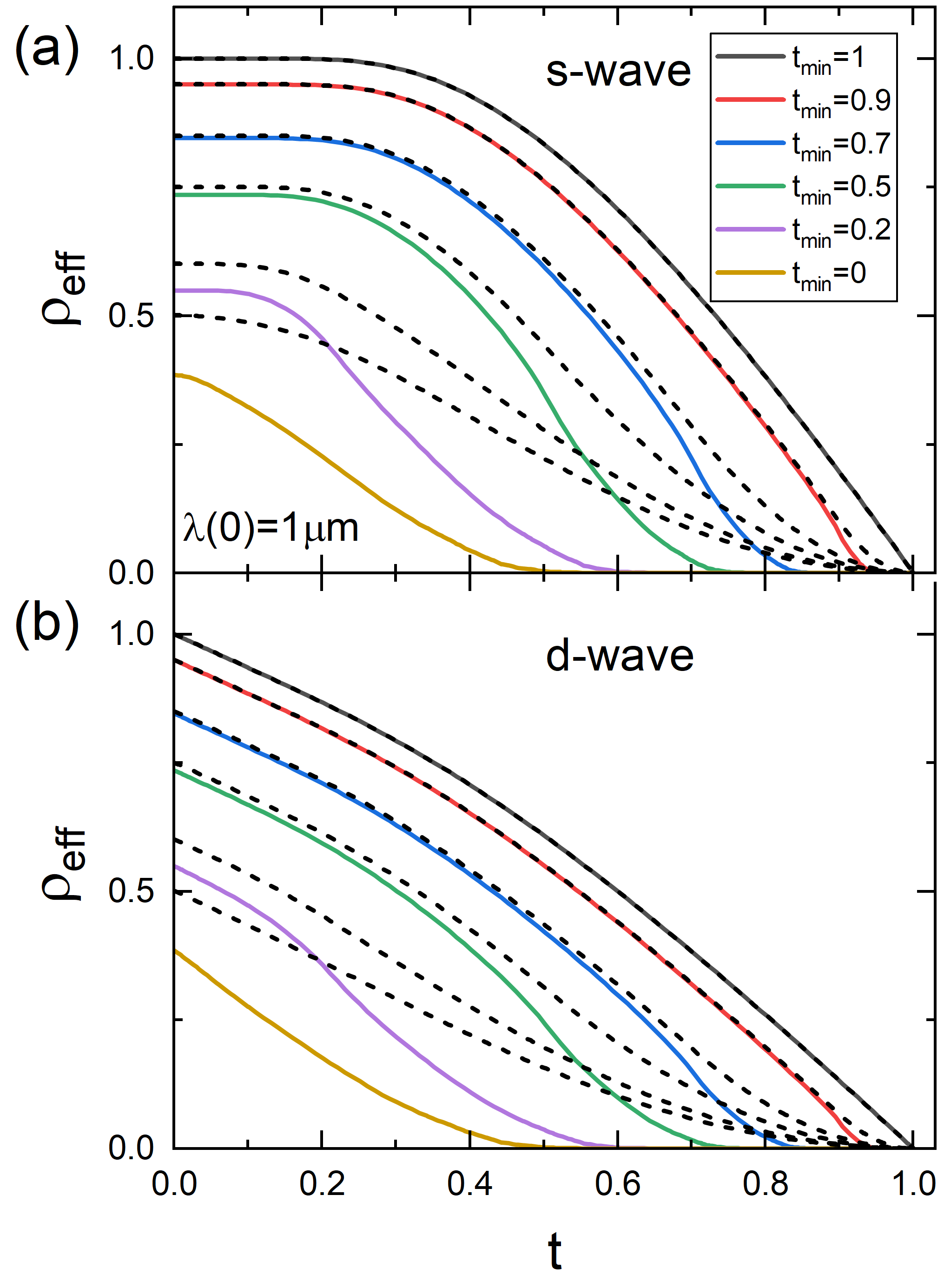}
\caption{Effective (measured) superfluid density $\reff=\Lambda_1(0)/\Leff$ of the mosaic (solid lines) and the area average (see text) of the local superfluid density (dashed lines) for (a) s-wave, and (b) d-wave domains. The label $\lambda(0)=1\,\mu$m denotes $\lambda_1(0)$, the zero-temperature penetration depth of the strongest domains ($T_c=T_{c,\max}$); the effective $\lambda_{\rm eff}(0)=\lambda_1(0)/\sqrt{\reff(0)}$ of the film is larger.}
\label{fig:rho}
\end{figure}

Figure~\ref{fig:rho} summarizes the results.
A narrow distribution, $\tmin=0.9$, gives an almost uniform curve with the onset of screening at $t_p=0.95$.
For broader distributions, the $\reff(t)$ curves shift to lower temperatures. As far as the low-temperature saturation is concerned, the results for the two pairing symmetries are very different. For s-wave, Fig.~\ref{fig:rho}(a), the flat region shrinks, yet remains finite and only disappears for $\tmin=0$, as expected. However, for d-wave domains, the low-temperature part is $t$-linear from the start. As shown in Fig.~\ref{fig:rho}(b), broadening the $T_c$ distribution also shifts the curves to lower temperatures, and the effective transition temperature follows the same trend as in the s-wave case, but the low-temperature behavior remains linear.

For contrast, the dashed lines in Fig.~\ref{fig:rho} show the area average of the local superfluid density, $\langle t_i\rho(t/t_i)\rangle$, which counts all superconducting domains regardless of connectivity and therefore extends to $t=1$. It is the upper bound on $\reff$. Therefore, such a simplified analysis of the inhomogeneous state, even with mosaic domains, does not reflect the experiment and is not correct.

\subsection{Magnetic susceptibility in the $T_c$--mosaic model}

\begin{figure*}[t]
\includegraphics[width=16cm]{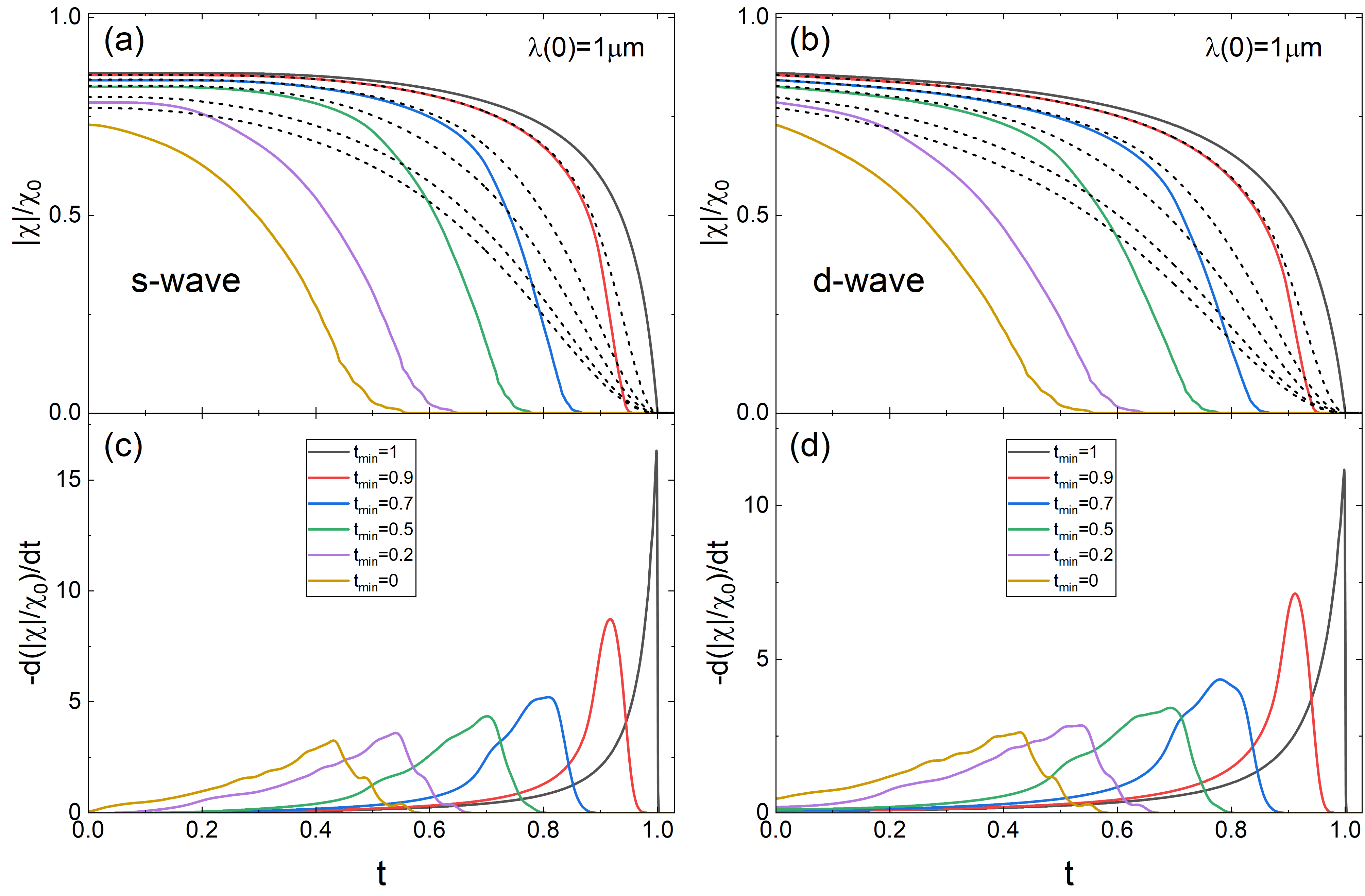}
\caption{(a,b) Magnetic susceptibility of a disc ($a=3$~mm, $d=20$~nm, $\lambda_1(0)=1\,\mu$m, labeled $\lambda(0)$ in the panels) for s-wave and d-wave domains. $\tmin=1$: uniform film. Dotted: uniform film with the area-averaged superfluid density [dashed lines in Fig.~\ref{fig:rho}], $\tmin=0.9$ to 0 from top to bottom. (c,d) $-d(|\chi|/\chi_0)/dt$, smoothed (except for $\tmin=1$) with a Gaussian of width $\delta t=0.01$ to suppress the steps produced by individual cluster connections in the finite box. A kink in $\chi(t)$ would appear as a step.}
\label{fig:chi}
\end{figure*}

Figure~\ref{fig:chi}(a,b) shows $|\chi(t)|/\chi_0$ over the full temperature range for s-wave (a) and d-wave (b) cases, together with its temperature derivative (c,d), respectively. As expected, $\chi=0$ above $t_p$, not only above $t=1$. (Note that above $t_p$, finite superconducting clusters retain some diamagnetic response, which is neglected in our homogenized-disc calculation.) Similarly to the superfluid density discussed above, the s-wave curves saturate at low temperatures, whereas the d-wave curves do not. Of course, in a real experiment, the minimum measurement temperature will limit the observation of this behavior. Furthermore, an important feature to compare with the experiment is the sharpness of the superconducting transition. For the uniform film ($\tmin=1$), $-d\chi/dt$ jumps to a finite value at $T_c$, whereas in the mosaic it rises continuously from zero at $t_p$, Fig.~\ref{fig:chi}(c,d) (see Sec.~\ref{sec:disc}), and the whole curve broadens more and more as $\tmin$ decreases.

\subsection{Low-temperature penetration depth}\label{sec:lowT}
\begin{figure}[tbh]
\includegraphics[width=8cm]{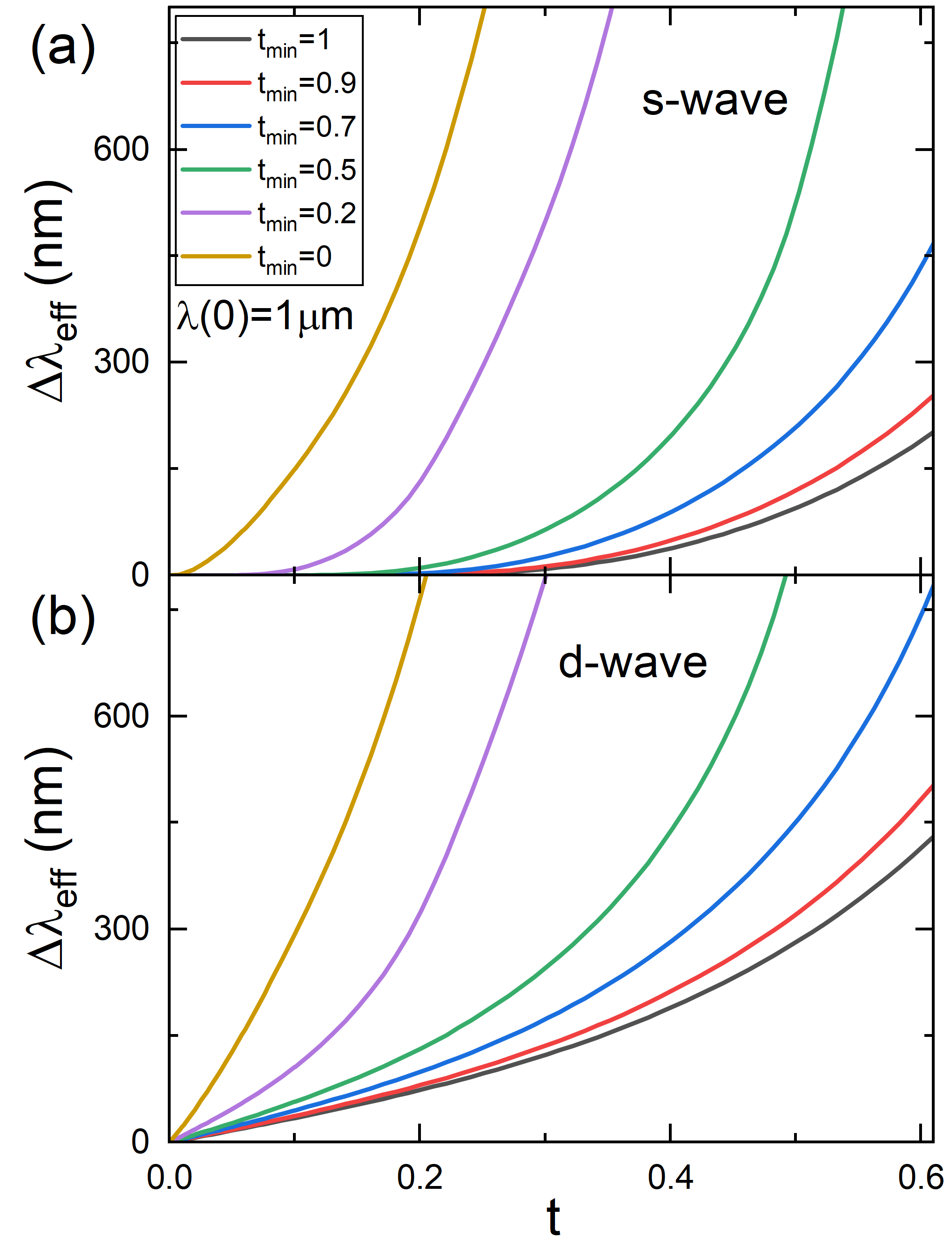}
\caption{Low-temperature part of $\Delta\lambda_{\rm eff}(t)=\lambda_{\rm eff}(t)-\lambda_{\rm eff}(0)$, in nm, for (a) s-wave and (b) d-wave domains, with $\lambda_1(0)=1\,\mu$m (labeled $\lambda(0)$ in the panel). For broad distributions, the curves leave the frame as $t\to t_p$, where $\lambda_{\rm eff}$ diverges.}
\label{fig:lowT}
\end{figure}

Figure~\ref{fig:lowT} shows $\Delta\lambda_{\rm eff}(t)=\lambda_{\rm eff}(t)-\lambda_{\rm eff}(0)$, the quantity a whole-sample measurement reports after calibration \cite{moirangthem2026b,prozorov2026films}.

For s-wave domains, Fig.~\ref{fig:lowT}(a), the behavior remains activated for every $\tmin>0$, with an activation energy set by the weakest domains, $\Delta(0)\approx1.76\,\kB T_{c,\min}$. The amplitude is larger for broader distributions, consistent with the reduction of the superfluid density. For $\tmin=0$, domains with $t_i\lesssim t$ keep losing their condensate down to $t\to0$. For this uniform distribution, the area average decreases as $\Delta\langle\rho\rangle\propto t^2$: the domains that lose their condensate have $t_i\lesssim t$, and their superfluid density is $\propto t_i$. For the film, the decrease is several times larger and the exponent somewhat smaller ($\approx1.8$), because the weak domains carry the largest supervelocity, $u>1$, and their loss enters Eq.~(\ref{eq:sens}) with the weight $u^2$. Power-law fits, $\Delta\lambda_{\rm eff}\propto t^n$, over windows within $t\le0.3$ give $n=1.6$--1.8 for $\tmin=0$ and $n>4$ for $\tmin=0.2$; the local exponent, $d\ln\Delta\lambda_{\rm eff}/d\ln t$, of the $\tmin=0$ curve never falls below 1.5.

For d-wave domains, Fig.~\ref{fig:lowT}(b), where $\lambda_i^{-2}(0)\propto t_i$, the superfluid density of every domain at low temperature is
\begin{equation}
t_i\,\rho_d(t/t_i)=t_i\Big(1-0.648\frac{t}{t_i}\Big)=t_i-0.648\,t.
\end{equation}
Every domain therefore loses superfluid density at the same absolute rate, independent of its $T_c$. By Eq.~(\ref{eq:sens}), the film loses it at
\begin{equation}
\frac{d\reff}{dt}\Big|_{t\to0}=-0.648\,\langle u^2\rangle,
\label{eq:dslope}
\end{equation}
where $\langle u^2\rangle\ge1$ is the mean-square supervelocity of the $t=0$ current pattern. For $\tmin=0$, $\langle u^2\rangle=1.48$. The relative slope $|d\ln\reff/dt|$ is then $0.648\times1.48/0.385=2.49$ instead of 0.648.
The initial linear slope of $\Delta\lambda_{\rm eff}/\lambda_{\rm eff}(0)$ is therefore enhanced by a factor $\langle u^2\rangle/\reff(0)\approx3.8$ relative to a uniform film, while the exponent is unchanged. Because $\lambda_{\rm eff}(0)=\lambda_1(0)/\sqrt{\reff(0)}=1.61\lambda_1(0)$, the absolute slope of $\Delta\lambda_{\rm eff}$ plotted in Fig.~\ref{fig:lowT}(b) is enhanced by $\approx6.2$ relative to a uniform film with $\lambda(0)=\lambda_1(0)$.
The computed initial slope, $d\reff/dt=-0.974$ between $t=0$ and $0.005$, agrees with the prediction $-0.648\times1.48=-0.961$ within the realization scatter of $\langle u^2\rangle$ ($\pm0.05$). For $\tmin=0.5$, $\langle u^2\rangle=1.020$. A large linear slope in a whole-sample measurement of an inhomogeneous nodal film therefore does not by itself imply a large nodal density of states. It may simply reflect a reduced effective superfluid density.

\subsection{Non-uniform $T_c$ distributions}\label{sec:nonuni}
The uniform distribution is not special. For an arbitrary normalized distribution of local critical temperatures, $P(t_i)$, the area-averaged superfluid density of s-wave domains decreases by
\begin{equation}
\begin{split}
\Delta\langle\rho\rangle(t)&=\int_0^1\!P(t_i)\,t_i\Big[1-\rho\Big(\frac{t}{t_i}\Big)\Big]dt_i\\
&=t\int_0^1\!P(t_i)\,f\Big(\frac{t}{t_i}\Big)dt_i,
\end{split}
\label{eq:dav}
\end{equation}
where $\rho(x)=0$ for $x\ge1$ and $f(x)=[1-\rho(x)]/x$. The function $f$ is bounded, $0\le f\le1$, exponentially small for $x\to0$, and is equal to $1/x$ for $x\ge1$. Hence, a domain with $t_i\gg t$ contributes much less than $t$, while the domains with $t_i\lesssim t$, which contribute at most $t$ each, occupy a fraction of the film that vanishes as $t\to0$. Therefore, $\Delta\langle\rho\rangle/t\to0$ for any distribution of local $T_c$ (formally, this follows from Lebesgue's dominated-convergence theorem: if functions converge pointwise and are all bounded by one integrable function, their integrals converge to the integral of the limit. Here, $P(t_i)f(t/t_i)\to0$ for every $t_i>0$ and is bounded by $P(t_i)$). On cooling, the area-averaged superfluid density always approaches its zero-temperature value with zero slope, $d\langle\rho\rangle/dt\to0$, and has no $T$-linear term.

For example, for a general power-law distribution, $P(t_i)=(1+\alpha)\,t_i^{\alpha}$ on $(0,1]$ with $\alpha>-1$, Eq.~(\ref{eq:dav}) gives exactly $\Delta\langle\rho\rangle\propto t^{\,n_{\rm av}}$ with $n_{\rm av}=2+\alpha$ for $t\ll1$, while the percolation temperature is still the median, $t_p=2^{-1/(1+\alpha)}$, and $\langle\rho\rangle(0)=(1+\alpha)/(2+\alpha)$. Eliminating $\alpha$,
\begin{equation}
n_{\rm av}=1+\frac{\ln2}{\ln(T_{c,\max}/T_p)}.
\label{eq:nav}
\end{equation}
The exponent approaches $n_{\rm av}=1$ only as the onset of screening is significantly suppressed: $n_{\rm av}=1.5$ for $T_{c,\max}/T_p=4$, 1.25 for 16, and 1.1 for 1024.
The log-uniform distribution, $P\propto1/t_i$ on $[\tmin,1]$, is the limit of the power-law distribution when $\alpha\to-1$. It gives $\Delta\langle\rho\rangle\propto t/\ln(1/\tmin)$ for $\tmin\ll t$, but $t_p=\sqrt{\tmin}$. For example, for $\tmin=0.01$, $t_p=0.1$: the film screens only below $0.1\,T_{c,\max}$, well below $T_{c,\max}/3$, the usual upper limit of the low-temperature range used to fit $\lambda(T)$.

\begin{table}[tbh]
\caption{Low-temperature penetration depth of mosaics with non-uniform $T_c$ distributions (s-wave unless marked d). $n_{\rm av}$: exponent of $\Delta\langle\rho\rangle=\langle\rho\rangle(0)-\langle\rho\rangle(t)$, Eq.~(\ref{eq:nav}) (fit for the log-uniform case). $n$: exponent of $\Delta\lambda_{\rm eff}\propto t^n$ from fits over the windows $0.05$--$0.3$, $0.1$--$0.5$ and $0.05$--$0.6$ of $t_p$ (range of the three values). Log-uniform is $\propto1/t_i$ on $[10^{-3},1]$.}
\label{tab:nonuni}
\begin{ruledtabular}
\begin{tabular}{lccccc}
$P(t_i)$ & $t_p$ & $\langle\rho\rangle(0)$ & $\reff(0)$ & $n_{\rm av}$ & $n$\\
\hline
uniform ($\tmin=0$) & 0.5 & 0.50 & 0.385 & 2 & 1.66--1.73\\
$\propto t_i^{-1/2}$ & 0.25 & 0.33 & 0.172 & 1.5 & 1.32--1.47\\
$\propto t_i^{-3/4}$ & 0.0625 & 0.20 & 0.042 & 1.25 & 1.11--1.23\\
log-uniform & 0.032 & 0.15 & 0.031 & 1.15 & 1.24--1.36\\
d: uniform ($\tmin=0$) & 0.5 & 0.50 & 0.385 & 1 & 1.19--1.34\\
d: $\propto t_i^{-1/2}$ & 0.25 & 0.33 & 0.172 & 1 & 1.09--1.26
\end{tabular}
\end{ruledtabular}
\end{table}

Of course, the average $n_{\rm av}$ is not what is measured. To see how the mosaic domains modify the behavior, we computed $\reff$ for different $P(t_i)$ distributions with the same geometry realizations at 37 temperatures up to $1.1\,t_p$, as summarized in Table~\ref{tab:nonuni}. For the s-wave power-law distributions, as for the uniform one, the film exponent lies below $n_{\rm av}$, because the weak domains that lose their condensate carry $u>1$. In units of its own onset of screening, the s-wave mosaic with $P\propto t_i^{-3/4}$ gives $\Delta\lambda_{\rm eff}\propto t^{1.1-1.2}$, as close to linear as the d-wave mosaic with a uniform distribution. However, in this case, the onset of screening lies 16 times below the highest local $T_{c,\max}$, half of the domains have their local $T_c$ between $T_p$ and $16\,T_p$, and $\reff(0)=0.04$, i.e., $\lambda_{\rm eff}(0)=4.9\,\lambda_1(0)$. The log-uniform case does not produce closer to linear behavior either, because its cutoff makes $\lambda_{\rm eff}$ activated again below $t\approx\tmin=0.03\,t_p$.

Still, the $n$ values in the range of 1.1-1.3 may look close to linear, especially with the always present noise in the experimental data. Therefore, the exponent alone cannot exclude an s-wave mosaic. The decisive quantity is the ratio $T_{c,\max}/T_p$ between the highest local $T_c$, seen by local probes (not only magnetic, but including, for example, tunneling spectroscopy), and the onset of macroscopic screening, $T_p$. Since these quite specialized measurements may not be available for every film studied, transport measurements can be also used by detecting the onset of the resistive transition, which will be visible even for non-percolating clusters. Importantly, the offset of the resistive transition, $R=0$, must be close to percolation threshold, $t_p$, and can be compared with the onset of macroscopic screening seen by the measurements of $\chi(T)$. For example, s-wave $n\le1.5$ requires $T_{c,\max}/T_p \gtrsim 4$, and $n \approx 1.2$ requires $T_{c,\max}/T_p \approx 16$, both easily resolvable.

\section{Discussion}\label{sec:disc}
\emph{Behavior near the onset of the transition.}

In the developed $T_c$--mosaic model, the superconducting transition seen in whole-sample or part-sample mutual-inductance measurements of $\chi(T)$, is a percolation transition of the global superfluid density. Its onset is at the median of the local $T_c$ distribution, $T_p$, not at the maximum, $T_{c,\max}$. A uniform film has $\rho\propto(T_c-T)$ near $T_c$. In weak screening, $|\chi|\approx(3\pi/32)(a/\Lambda)\chi_0\propto\rho$, so $\chi$ starts linearly from $T_c$ and $d\chi/dT$ jumps at $T_c$, which is a kink in $\chi(T)$. The mosaic, on the other hand, has $\reff\propto(t_p-t)^{\mu}$ in the critical region, with $\mu>1$ as for the conductivity of two-dimensional percolating networks ($\mu\approx1.3$ asymptotically \cite{efros1976,stauffer1994}; effective exponents fitted over a wider range near $t_p$ are larger), so $d\chi/dT$ rises continuously from zero as shown in Fig.~\ref{fig:chi}(c,d). The onset is a soft kink in the critical region and a rounded foot outside it. Below the onset, the curve is broader than that of the uniform film ($\tmin=1$ in Fig.~\ref{fig:chi}), because superconductivity strengthens gradually as more domains join in and as each domain's own $\rho$ grows. For $\tmin=0$ the 50\% point of $\chi$ is at $t_{50}=0.36$ (s-wave) and 0.33 (d-wave) for an onset at $t_p=0.5$.
Regarding the steepness of $\chi(T)$ below the transition, because $\Leff$ crosses $a$ only when $\reff$ drops to 0.033, $\chi(T)$ rises steeply below the onset and then saturates. The strong-screening plateau hides most of the low-temperature variation of $\reff$, which is visible only through the logarithm in $1-F$. Above the percolation transition, there is no measurable magnetic susceptibility (neglecting diamagnetism of non-percolating clusters which do not have large-loop currents). Yet, superconductivity starting at $T_{c,\max}$ can still be clearly visible to local probes when they focus on the domains with maximum screening. The same transition can be detected as the onset in resistivity.

Therefore, analysis of the measured $\chi(T)$ and resistivity near the apparent transition, as well as comparison with local probes (if available) can be used to determine parameters of the $T_c$--mosaic.

\emph{Distinguishing effect of inhomogeneity from genuine nodes.}

For s-wave domains, inhomogeneity alone may produce a $\lambda(T)$ that does not saturate only if a finite fraction of the film has a local $T_c$ near zero, $\tmin\to0$. However, the same sample will show a greatly reduced onset of the transition, $t_p\le0.5$ for the uniform distribution and even lower for distributions weighted toward $T_c\to0$ (Sec.~\ref{sec:nonuni}), and a strongly suppressed $\reff(0)$.

Line nodes in uniformly scaled d-wave domains produce a linear term in $\lambda(T)$ whose relative slope is enhanced by $\langle u^2\rangle/\reff(0)$, Eq.~(\ref{eq:dslope}).
Local probes with resolution better than $\ell$ measure single-domain $\rho(T)$ and are insensitive to the network. Comparing the local and whole-sample $\lambda(T)$ on the same film can therefore separate the two contributions.

\emph{Assumptions and limitations of the $T_c$--mosaic model.}

The developed $T_c$--mosaic  model has limitations that follow from the assumptions made: (1) The calculation is done in the mean-field, local London approximation. This requires $\xi\ll\ell$ and excludes nonlocal effects in clean nodal superconductors. (2) The domains are independent, with uncorrelated $T_c$. (3) There is no coupling through normal domains. Josephson or proximity coupling through them would connect clusters above $t_p$ and increase the transition onset. For domains of size $\ell=20\,\mu$m, much larger than the proximity length ($\lesssim0.1\,\mu$m), this coupling is negligible. (4) Phase fluctuations are not included. Near $t_p$, the superfluid density of the mosaic network is small and vanishes continuously at $t_p$. Therefore, the Berezinskii--Kosterlitz--Thouless transition, at which the Pearl length reaches the universal value $\Lambda(T_{\rm BKT})=\Phi_0^2/(4\pi\mu_0 k_BT_{\rm BKT})$ \cite{berezinskii1972,kosterlitz1973,nelson1977,beasley1979,hebard1980}, always lies below $t_p$, and vortex--antivortex unbinding would further reduce the onset of screening.

\section{Conclusions}

Within the presented $T_c$--mosaic model, a superconducting film made entirely of domains with different critical temperatures screens a perpendicular field only below the percolation threshold temperature, $T_p$, of its superconducting domains, which is the median local $T_c$. Above $T_p$, only very weak residual macroscopic diamagnetic response of individual non-percolating clusters remains and is neglected in this model.

Fully gapped s-wave domains with a positive lower cutoff, $\tmin>0$, retain an asymptotically (at $t\to0$) activated penetration depth, but its temperature interval shrinks with decreasing $t_{\min}$. Extending the $T_c$ distribution to zero, $t_{\min}\to0$, removes this activated behavior while substantially suppressing the superfluid density and an apparent transition temperature, $T_p$.
For uniform distributions of $T_c$, the fully gapped mosaic does not yield a $T$-linear behavior over any temperature window: $\Delta\lambda_{\rm eff}\propto t^n$ with $n=1.6$--1.8 for $\tmin=0$ and $n>4$ for $\tmin=0.2$.
For any general distribution, the area-averaged superfluid density approaches its zero-temperature value with zero slope, without a $T$-linear term, and for power-law distributions the exponent of $\langle\rho\rangle(0)-\langle\rho\rangle(t)$ is tied to the onset of screening, $n_{\rm av}=1+\ln2/\ln(T_{c,\max}/T_p)$.
The behavior comes closer to linear than the area average, but only in units of $T_p$ and only when $T_p$ lies far below the highest local $T_c$: $n\le1.5$ requires $T_{c,\max}/T_p\gtrsim4$, and $n\approx1.2$ requires $T_{c,\max}/T_p\approx16$ and $\reff(0)\approx0.04$ (Table~\ref{tab:nonuni}).

Therefore, a $T$-linear $\lambda(T)$ measured on a film whose onset of macroscopic screening is within a factor of a few of the highest local $T_c$, as seen by local probes or by the onset of the resistive transition, cannot be explained by $T_c$ mosaicity of a fully gapped superconductor.

Nodal domains retain a linear low-temperature term (or whatever power law the individual domains have, set by impurity scattering) whose relative slope is enhanced by inhomogeneity.

These conclusions apply to the analysis of the magnetic susceptibility of the whole sample measured by global AC susceptibility and to the mutual-inductance technique, where the excitation field varies on millimeter scales, much larger than $\ell$, so the same homogenized $\Leff$ applies.

\begin{acknowledgments}
I thank Bicky Moirangthem for discussions. Claude AI was used to develop Python code for numerical calculations, which was then manually validated.
This work was supported by the U.S. Department of Energy, Office of
Science, National Quantum Information Science Research Centers, Superconducting
Quantum Materials and Systems Center (SQMS), under Contract No. 89243024CSC000002.
Ames National Laboratory is supported by the U.S. Department of Energy (DOE),
Office of Science, Basic Energy Sciences (BES), Materials Science \& Engineering
Division (MSED) and is operated by Iowa State University for the U.S. DOE under
contract DE-AC02-07CH11358.
\end{acknowledgments}

%

\end{document}